\documentclass[10pt,journal,comsoc]{IEEEtran}
\usepackage{graphicx}
\usepackage{array}
\usepackage{booktabs}
\usepackage{tabularx}
\usepackage{cite}
\usepackage[table]{xcolor}
\usepackage{tikz}
\usetikzlibrary{positioning,fit,backgrounds,arrows.meta}
\usepackage[hidelinks]{hyperref}
\hypersetup{colorlinks=true,linkcolor=blue,citecolor=blue,urlcolor=black}

\title{Agent-Native Wireless Communications: Architecture, Opportunities, and the Road Ahead}

\author{Yuanwei Liu, Xu Gan, Zhaolin Wang, Shan Shan, Zongyao Zhao, and Zhiguo Ding

\thanks{Y. Liu, X. Gan, Z. Wang, and Z. Zhao are with the Department of Electrical and Computer Engineering, The University of Hong Kong, Hong Kong (e-mails: \{yuanwei, eee.ganxu, zhaolin.wang, zongyao\}@hku.hk).}
\thanks{S. Shan is with Beijing University of Posts and Telecommunications, Beijing 100876, China (e-mail: shan.shan@bupt.edu.cn).}
\thanks{Z. Ding is with Nanyang Technological University, Singapore 639798 (e-mail: zhiguo.ding@ntu.edu.sg).}}

\begin{document}
\maketitle

\begin{abstract}
Future wireless networks are moving toward autonomous service operation, where network control and resource management need to respond to time-varying radio conditions and evolving service objectives. To address this shift, this article develops an agent-native wireless communication framework that characterizes the interplay between agent intelligence and communication systems. In this framework, the coupling is organized around \emph{agents for communications} and \emph{communications for agents}. For agent-native operation, the architecture is organized around deployable computing infrastructure, programmable open radio access network (O-RAN) software, and controllable communication interfaces. Based on this architecture, \emph{agents for communications} addresses the use of agents in communication-system design and operation, including agent-generated communication software and agent-driven adaptive wireless optimization. On the other side, \emph{communications for agents} addresses wireless service support for agent operation, including network-supported single-agent loops and network-assisted multi-agent coordination. Finally, it outlines promising research directions for measurable, safe, and interoperable deployment of agent-native wireless communications.
\end{abstract}

\section{Introduction}
Conventional wireless networks have been engineered through protocol functions and radio resource management across several generations of cellular systems~\cite{shafi2017tutorial}. This design principle decomposes network operation into control and optimization problems with well-defined objectives, constraints, and interfaces. As deployments become denser~\cite{kamel2016ultradense} and services more heterogeneous~\cite{afolabi2018slicing}, however, communication functions are increasingly coupled across layers and domains. Network operation is therefore difficult to capture by these static models, manually configured policies, or offline optimization alone. To address this limitation, artificial intelligence (AI)-native wireless communications embed learning, inference, and optimization into communication functions~\cite{shafin2020ai,lin2025ainative}. This enables data-driven prediction, adaptation, and resource optimization under complex operating conditions~\cite{wang2020thirty}.

\begin{figure*}[t]
\centering
\includegraphics[width=0.92\textwidth]{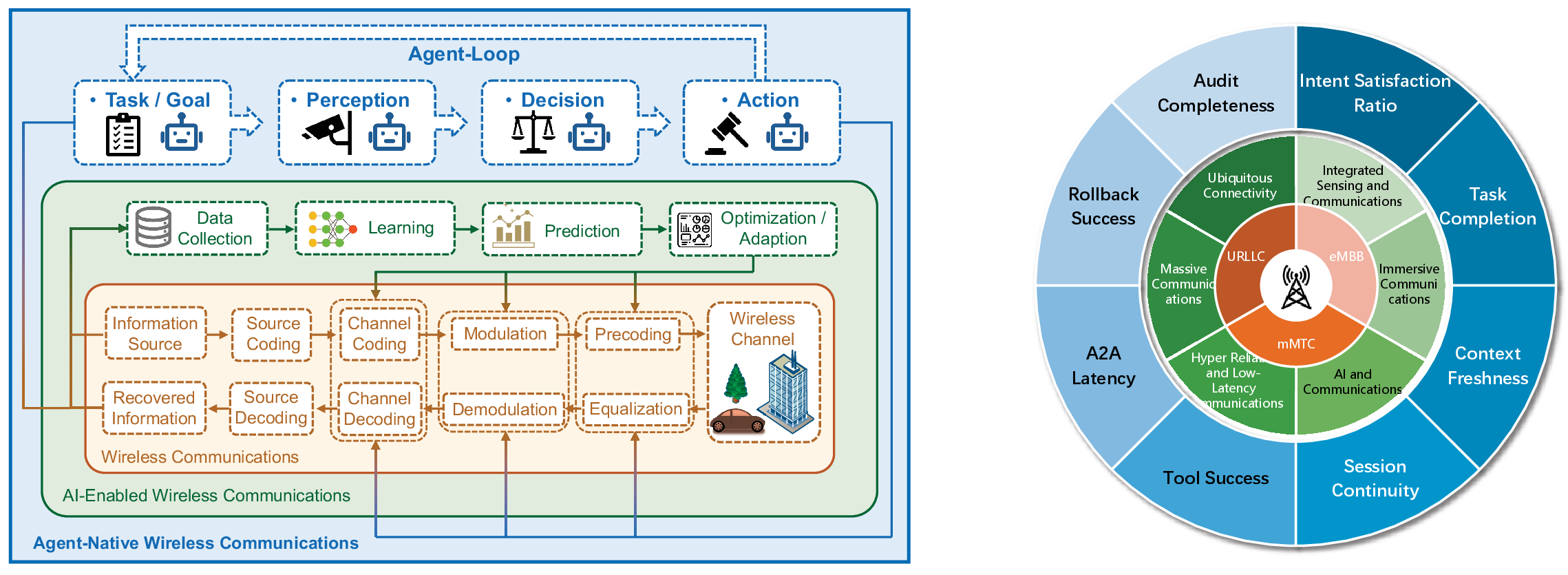}
\caption{Evolution of signal-processing principles and performance focus across conventional, AI-native, and agent-native wireless communications.}
\label{fig:compare}
\end{figure*}

However, future wireless networks require more than accurate prediction or a single optimized control output. AI-native wireless communications improve function-level intelligence, but this capability does not fully capture the service operation process over time~\cite{benzaid2020zerotouch,wang2020aiwireless}. Thus, such a process needs to maintain service intent, communication context, authorized tools, actions, and feedback in a closed loop. Agent-native wireless communications address this gap by introducing agents as intelligent software systems that support closed-loop operations. Fig.~\ref{fig:compare} summarizes this evolution along the signal-processing principle and the performance focus. The signal-processing principle moves from fixed model-based processing to AI-native function adaptation and then to agent-native closed-loop operation. The performance focus expands from link-level metrics to adaptive communication gains and verified loop-level effects. This loop-level view separates agent-native wireless communications into two complementary roles:

\begin{itemize}
    \item \textbf{Agent} refers to an intelligent software system that uses a large language model (LLM) as its reasoning core and supports autonomous intent understanding, action planning, task memory, and tool use. In wireless communications, an agent can interpret service objectives and communication context, and invoke authorized tools or network functions for communication-system operation~\cite{khaldi2026agenticdt,saleh2026agentictinyml}.
    \item \textbf{Agent loop} refers to the perception-decision-action-feedback process through which an agent operates in a wireless system. The loop connects context acquisition, decision making, authorized action execution, and outcome feedback, so that each communication action is evaluated against the context in which it was taken and then used to update subsequent decisions.
\end{itemize}

These two concepts define the basic unit of agent-native wireless communications. The agent provides intent interpretation and authorized action, while the agent loop connects perception, decision, action, and feedback. Wireless communications either embed this loop inside the network to improve design and operation or support it as a service for agent operation. This distinction leads to two coupled design views for agent-native wireless communications.

\begin{itemize}
    \item \textbf{Agents for communications} places agents and agent loops inside communication-system design and operation. The focus is on using service intent, network context, authorized tools, and feedback to generate communication software, configure protocol functions, and support adaptive wireless optimization.
    \item \textbf{Communications for agents} treats agent loops as wireless service processes that need network support. The focus is on timely and reliable support for context delivery, action transport, feedback association, stateful edge access, and multi-agent coordination.
\end{itemize}

\begin{table*}[t]
\centering
\caption{Coupling between agents and communications in agent-native wireless systems.}
\label{tab:two_views}
\renewcommand{\arraystretch}{1.2}
\setlength{\tabcolsep}{4pt}
\small
\rowcolors{2}{gray!12}{white}
\begin{tabularx}{\textwidth}{>{\raggedright\arraybackslash}p{0.14\textwidth} >{\raggedright\arraybackslash}X >{\raggedright\arraybackslash}X >{\raggedright\arraybackslash}X}
\toprule
\textbf{Design aspect} &
\textbf{Agents for communications} &
\textbf{Communications for agents} &
\textbf{Main implication} \\
\midrule
Task goal &
Communication-function optimization &
Agent-loop service continuity &
Joint agent--communication objective \\

Loop operation &
Intent-to-action network control &
Context-to-decision message delivery &
Closed-loop agent--communication coupling \\

Time scale &
RAN control time scale &
Freshness and deadline window &
Timing-aware loop operation \\

Coordination scope &
Shared radio-resource control &
Shared task-state consistency &
Resource and state coordination \\

Interface support &
Authorized control interface &
State-associated service interface &
Loop-level interface semantics \\

Evaluation focus &
Communication-effect verification &
Loop-continuity evaluation &
Closed-loop effectiveness metrics \\
\bottomrule
\end{tabularx}
\rowcolors{2}{}{}
\end{table*}

These two views make agent-native wireless communications a joint design problem of agent capability and communication support. Table~\ref{tab:two_views} places this joint design in wireless-system terms and relates the two views through common design aspects. The table highlights the agent loop as a common design unit for network control and service delivery. From this perspective, agents and communications are coupled through the same closed-loop process. The value of an agent action comes from its verified communication effect, while the value of communication support comes from preserving the state needed by the next agent decision.

The rest of this article follows this question from architecture to operation. It first explains the role of computing infrastructure, open radio access network (O-RAN) programmability, and open interfaces in deploying agent loops in wireless systems. It then studies agents inside communication systems through software generation and adaptive optimization, and examines wireless support for agent services through single-agent loops, multi-agent coordination, and agent-loop communication technologies. The conclusion summarizes deployment issues in measurement, safety, and interoperability.

\section{The Architecture of Agent-Native Wireless Communications}
This section develops the architectural basis for deploying agent loops in wireless networks. An agent-native architecture relies on three connected functions. First, computing infrastructure supports agent reasoning, validation, and state continuity. Second, O-RAN exposes RAN state and programmable control functions to agents. Third, open and programmable interfaces translate agent decisions into authorized network-configuration actions and return execution evidence to the loop. These components determine whether agents can interpret wireless operating state, act on communication functions, and maintain loop state across network changes.

\begin{figure*}[t]
\centering
\includegraphics[width=0.92\textwidth]{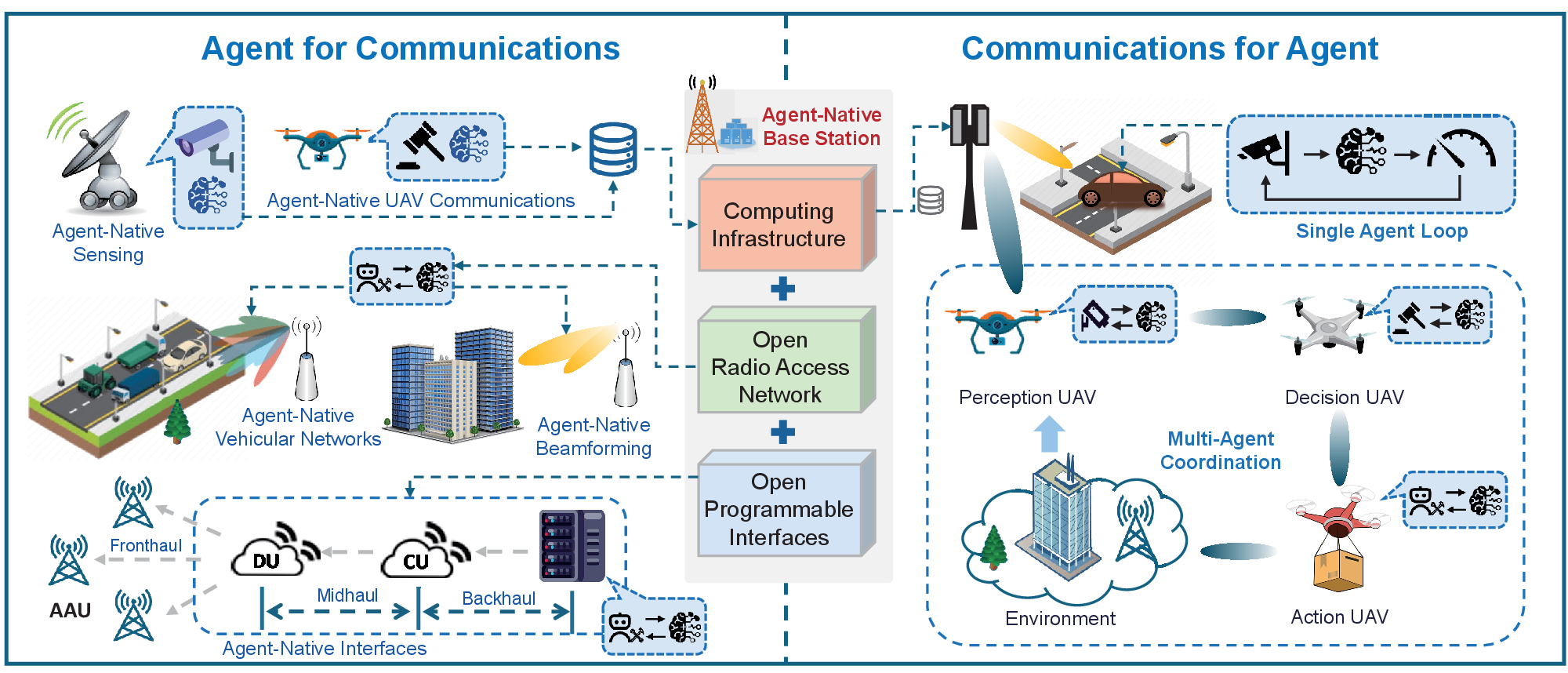}
\caption{Architectural support for agents for communications and communications for agents through computing infrastructure, O-RAN programmability, and open interfaces.}
\label{fig:architecture}
\end{figure*}

Fig.~\ref{fig:architecture} links the three architectural components to the two design views. Computing infrastructure provides the placement of agent reasoning and loop state across devices, edge nodes, and cloud resources. O-RAN builds on this placement by exposing RAN measurements and programmable control functions that agents can observe and invoke. Open interfaces turn the resulting decisions into authorized network actions and return execution evidence to the loop. With these components, \emph{agents for communications} can operate as controlled network entities, and \emph{communications for agents} can preserve agent-loop state across radio access and edge changes.

\subsection{Computing Infrastructure}
Computing infrastructure provides the hardware platform that runs the ``brain'' of agents in wireless systems~\cite{huang2025llmnetworking}. Its role is not only to supply processing resources for LLM reasoning, but also to place reasoning, memory, and validation close to the wireless state on which an action depends. The architectural issue therefore shifts from computing-resource allocation alone to the placement of agent intelligence and state across devices, edge nodes, RAN controllers, and clouds.

\begin{itemize}
    \item \textbf{Network-side agent reasoning.} An agent can change a protocol function or resource-control policy only when reasoning and validation finish within the control time scale of that function. Computing resources on the device, edge, or RAN side therefore become part of the latency budget for agent-driven network operation.
    \item \textbf{Service-side agent-loop continuity.} An agent loop may keep packet connectivity after handover or edge migration while losing the state needed for its next decision. Computing infrastructure provides state placement and migration support beyond packet forwarding.
\end{itemize}

Computing hardware therefore serves both as the reasoning platform for network-side agents and as the state-continuity platform for service-side agent loops. This calls for a distributed computing design in which latency-sensitive reasoning is placed close to radio operation, while larger-scale reasoning, validation, and state synchronization are supported by edge, regional, and cloud resources. In such a design, the location and scheduling of computing infrastructure resources become core architectural problems for agent-native wireless systems.

\subsection{O-RAN}
O-RAN provides the software platform that makes radio access network (RAN) operation accessible to agents~\cite{polese2023oran}. It turns radio measurements, control policies, and network functions into software objects that an agent can observe, reason about, and invoke. The architectural issue is to align agent decisions with the RAN functions, control authority, and feedback paths available at each control time scale.

\begin{itemize}
    \item \textbf{Network-side RAN programmability.} A network-side agent acts on RAN operation through software functions with clear control scope and time scale. O-RAN provides these functions through programmable RAN control, controllers, and service management mechanisms.
    \item \textbf{Service-side RAN-state feedback.} For an agent loop supported by wireless access, RAN conditions affect whether context remains fresh, actions arrive within their execution windows, and feedback can be associated with the right operating state. O-RAN can report these conditions to support timing, continuity, and feedback association for agent services.
\end{itemize}

In this architecture, O-RAN supports network-side agent actions through programmable RAN functions and service-side agent loops through RAN-state feedback. Agent-native operation depends on assigning each agent decision to the RAN control function and control time scale where it can be executed and verified. This requires mechanisms for translating agent intent into RAN control objectives, selecting the proper control time scale, reporting standardized RAN state, and returning execution feedback for agent actions.

\subsection{Open and Programmable Interfaces}
Open and programmable interfaces provide the control entry point through which agent reasoning becomes network configuration and control. Through these interfaces, an agent can adjust communication functions such as scheduling, mobility control, slicing, routing, and edge-service binding under explicit authorization. The network response, including execution status and measured performance change, is then returned as feedback for the next agent decision.

\begin{itemize}
    \item \textbf{Agent-driven network configuration.} Agents need interfaces that translate a decision into a bounded network-configuration action, such as updating scheduling, mobility, slicing, or edge-service control. The interface specifies the authority scope and control target so that the decision can be executed as a network operation.
    \item \textbf{Closed-loop execution verification.} After a configuration action is applied, the interface should return execution evidence that links the network response to the issued action. This allows the agent loop to verify the communication effect and trigger recovery when the response deviates from the expected behavior.
\end{itemize}

Programmable interfaces make agent-native operation executable by connecting agent decisions with controlled network actuation and measurable execution feedback~\cite{bonati2021oranlearning}. When agents modify communication functions, the interface should define the control scope, record the affected network state, and return execution evidence for the next loop iteration.
A practical interface layer treats each agent action as a controlled network operation rather than a direct configuration command. Observation, actuation, feedback, and recovery are then tied to one action record, keeping agent-driven control authorized and verifiable during wireless operation.

\section{Agents for Communications}
This section studies \emph{agents for communications} through the role of agent loops inside communication-system design and operation. In communication software development, the agent loop turns communication design objectives into verifiable software implementations. In wireless network operation, the agent loop adapts communication parameters to network dynamics and coordinates decisions within and across protocol layers. In both cases, the agent loop keeps each generated artifact or control action tied to the corresponding specification, constraint, interface, and feedback.

\subsection{Agent-Generated Communication Software}
In communication software development, agents can connect system-level design objectives with software implementation and verification. Specifically, a communication agent loop can generate software artifacts such as protocol logic, control algorithms, implementation code, test cases, and simulation scripts. This changes the workflow from separated manual coding and testing to an interactive process in which generated software is revised according to verification results.

\begin{itemize}
    \item \textbf{Specification-to-implementation generation.} For a new communication function, an agent can generate executable software and validation tests from the specification. The following simulation and test results guide the agent to refine the implementation until its protocol behavior satisfies the design objective.
    \item \textbf{Programmable-network software restructuring.} Existing communication implementations are often written for fixed procedures and closed execution environments. Agent-assisted restructuring can turn such implementations into software modules with explicit control inputs and interface requirements, so that they can be configured, tested, and integrated in programmable networks.
\end{itemize}

These two roles extend agent support from the creation of new communication software to the modernization of existing implementations for programmable network operation. Here, the agent loop becomes a development loop: the agent reads specifications and verification results, decides which software component to generate or revise, updates the implementation, and uses simulation or test feedback for the next revision. Its role is to keep the generated software aligned with communication constraints throughout this process. For instance, a MAC-layer scheduling module has to allocate resource blocks for each transmission time interval according to queue state, hybrid automatic repeat request feedback, and modulation-and-coding constraints; otherwise, the generated scheduler may pass software tests while producing invalid radio-resource decisions. To support this revision process, a controlled workflow is needed in which the agent keeps each generated or restructured artifact linked to its communication specification, verification evidence, and target interface before deployment.

\subsection{Agent-Driven Communication Optimization}
In wireless network operation, agents adapt communication parameters as radio conditions, traffic load, interference, and mobility vary. The agent loop observes the current network condition, selects a bounded parameter update, applies it through authorized network functions, and checks the measured response before the next update. This closed-loop agent operation extends function-level AI from producing an optimized output to maintaining the relation among network state, selected action, and observed response over repeated control cycles. When multiple agents operate in the same wireless system, optimization has to be coordinated within one protocol layer and across different protocol layers~\cite{luong2019drl}.

\begin{itemize}
    \item \textbf{Intra-layer agent coordination.} Within one protocol layer, local control decisions are distributed but still coupled through shared radio resources. Specialized agents can optimize different local decisions in parallel, while intra-layer coordination keeps their parameter updates consistent with the same resource state.
    \item \textbf{Cross-layer agent coordination.} A local update at one layer may change the feasible decisions and performance limits of another layer. Cross-layer coordination places service objectives, resource allocation, and link-state adaptation in one decision chain, so that layer-specific updates translate into end-to-end communication performance improvement.
\end{itemize}

\begin{figure*}[!t]
\centering
\includegraphics[width=0.7\textwidth]{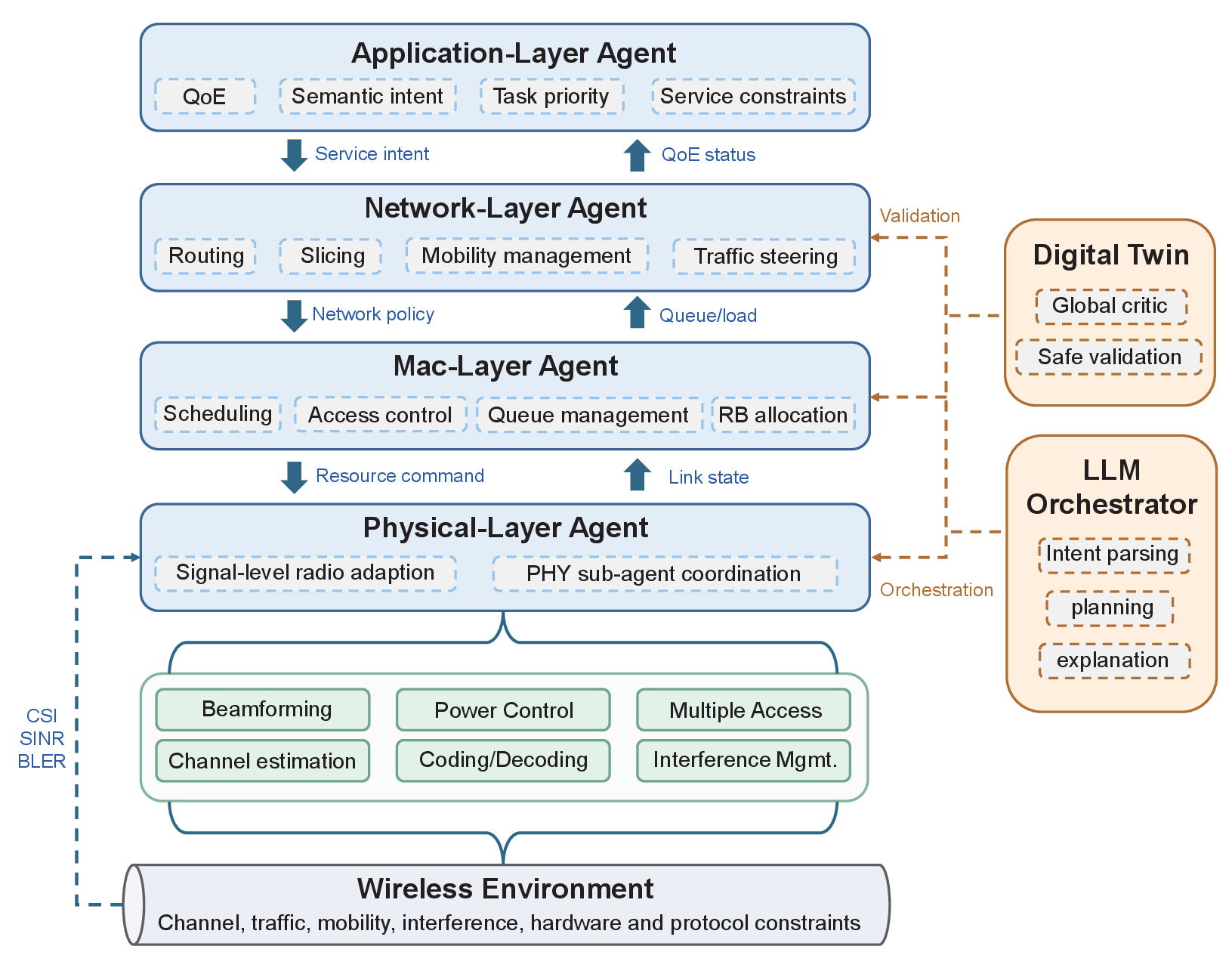}
\caption{Four-layer hierarchical multi-agent framework for autonomous wireless optimization.}
\label{fig:framework}
\end{figure*}

These two coordination modes correspond to different parts of the same layered control structure. Intra-layer coordination appears among specialized agents that share resources within one layer, while cross-layer coordination appears along the application, network, MAC, and PHY control chain. Fig.~\ref{fig:framework} illustrates how multi-agent optimization is organized across these communication functions. At the PHY layer, specialized agents tune local radio functions over the same wireless environment, so their actions have to be coordinated as one local resource decision. Across layers, service-level targets are translated into network policy, MAC scheduling, and PHY radio adaptation, while bottom-up feedback indicates whether the bottleneck comes from routing or slicing, queue buildup, link quality, or radio interference. The digital twin evaluates candidate control actions before execution, and the LLM orchestrator translates service intent into layer-specific control objectives for the agents involved in the update.

\begin{figure}[!t]
\centering
\includegraphics[width=1\linewidth]{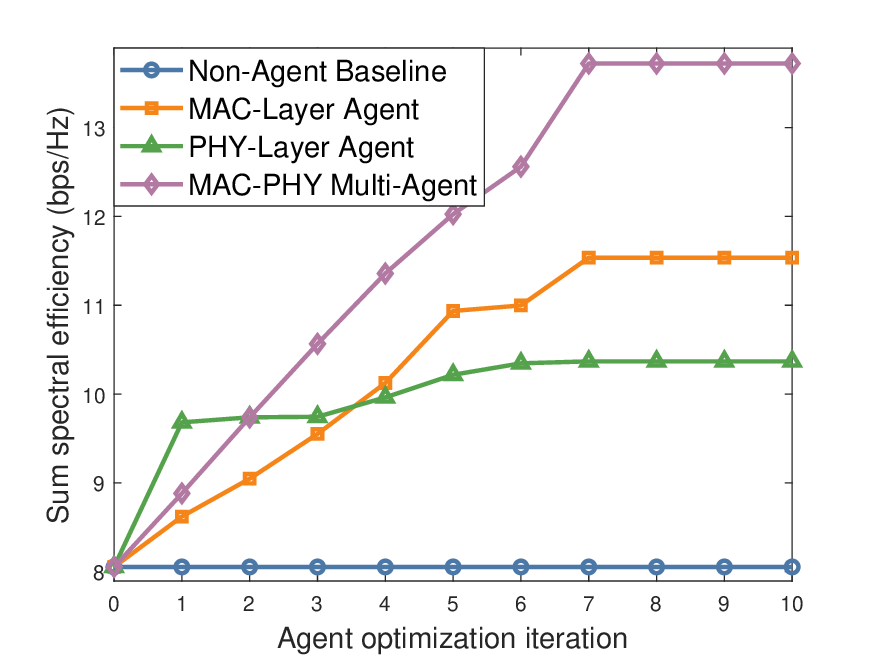}
\caption{Illustrative MAC-PHY multi-agent optimization.}
\label{fig:multi-agent-simulation}
\end{figure}

\textit{Use case: MAC-PHY optimization.} We consider joint MAC-layer access control and PHY-layer beamforming in a single-cell line-of-sight multi-user multiple-input multiple-output (MIMO) system as an example. The base station selects active users in each transmission interval. The MAC-layer agent selects the user group according to channel strength and spatial correlation, while the PHY-layer agent refines the beamforming vectors with a regularized zero-forcing initialization. The two actions are coupled because user selection determines the spatial interference that beamforming has to suppress. Fig.~\ref{fig:multi-agent-simulation} shows that single-layer agents already improve spectral efficiency, but the joint MAC-PHY agents achieve the largest gain by optimizing user grouping and beamforming as one coupled communication decision. The staged convergence also indicates that most of the gain can be obtained within a small number of agent iterations, which is important when coordination consumes signaling and computation resources.

\section{Communications for Agents}
This section studies \emph{communications for agents} through the wireless support needed to sustain agent loops during service operation. Conventional wireless services are commonly specified through user connectivity, application sessions, and quality-of-service (QoS) flows. Agent services introduce loop-level requirements because message value depends on the state of the agent loop. In particular, observations require freshness, commands require execution-window reliability, and feedback requires association with the action and context being updated~\cite{kountouris2021semantics}. The discussion therefore uses the number of agents and the number of loops as the organizing dimensions. Network-supported single-agent loops focus on freshness, deadline-constrained reliability, and service continuity for one agent. Network-assisted multi-agent coordination focuses on shared context, coordination signaling, and group-level consistency among agents.

\subsection{Network-Supported Single-Agent Loops}
Network-supported single-agent loops focus on wireless support for an agent whose decisions depend on messages delivered through the network. The service object extends from terminal connectivity to the continuity of the agent loop, because message usefulness depends on its validity. This loop-oriented service view is reflected in two examples. In connected robot control, one agent loop depends on sensing updates, control commands, and execution feedback within each control cycle. In an edge-assisted mobile agent service, the same agent can combine a fast on-device loop with slower edge inference, planning, and memory update loops. These examples lead to two communication cases: loop-level delivery for a single control loop and state association across multiple loops of the same agent.

\begin{itemize}
    \item \textbf{Single-agent single-loop.} In a single agent loop, wireless communication carries the context, command, and feedback used by the same decision cycle. The value of these messages depends on their role in the loop: a context update supports the current perception, a command triggers the corresponding action, and feedback closes that action for the next decision. Radio scheduling and QoS-flow mapping should therefore follow the validity of each loop message, so that wireless delivery preserves the timing and causality of the agent loop.
    \item \textbf{Single-agent multi-loop.} A single agent may run a fast local-control loop together with edge-side inference, planning, and memory-update loops. These loops may use different QoS flows or radio bearers, but their messages still update the same agent state. Wireless support should coordinate QoS adaptation, radio scheduling, context transfer, and edge continuity so that local actions and edge-side updates remain consistent for the next decision.
\end{itemize}

These two cases show that wireless communication for a single agent should treat packets as updates of the agent state. Each message needs to remain associated with its loop identity, state version, related action, and residual validity time. Such loop-state information allows the network to prioritize decision-critical messages, discard outdated context, and migrate relevant state during handover or edge migration. This state-continuity service defines the communication role of network-supported single-agent loops.

\subsection{Network-Assisted Multi-Agent Coordination}
Network-assisted multi-agent coordination extends the single-agent objective from agent-state continuity to shared-state consistency among agents. The communication service is no longer only to deliver messages to a group, but to help multiple agents maintain a consistent understanding of the same task state. Two examples illustrate this setting. In cooperative robot or vehicle control, several agents may contribute observations, actions, and feedback to one collaborative loop. In distributed edge intelligence, each agent may keep a local loop while also participating in a group coordination loop. These two examples motivate the following distinction between a shared loop across agents and coupled local loops with group coordination.

\begin{itemize}
    \item \textbf{Multi-agent single-loop.} A shared multi-agent loop involves several agents that jointly maintain one task state through observations, decisions, and feedback. Link-level reliability alone cannot ensure coordination when agents act on different versions or ages of the shared context. Wireless communications therefore need to align shared-context updates across agents, with groupcast, sidelink delivery, and edge aggregation providing possible support.
    \item \textbf{Multi-agent multi-loop.} Multi-agent multi-loop operation combines local autonomy with group coordination. Each agent updates its own local loop, while a coordination loop exchanges the shared state needed for joint decisions. Since these loops run on different time scales, the network has to control synchronization interval, update granularity, and coordination signaling to avoid excessive status exchange while keeping agents coordinated.
\end{itemize}

The common view is that multi-agent coordination makes wireless communication part of shared-state maintenance. Therefore, each coordination message can be associated with a compact coordination-state tag that records the sending agent, the shared-state version being updated, the related proposal or commitment, and the validity window of the update. This tagged state information supports group delivery, reliable commitment synchronization, and suppression of obsolete shared-state updates, forming the communication basis of network-assisted multi-agent coordination.

\subsection{Enabling Technologies for Agent-Loop Communications}
The four agent-loop settings above point to three communication capabilities for agent services. First, radio-edge service continuity keeps an agent loop and its state connected across wireless access, handover, and edge relocation. Second, wireless group-state synchronization aligns shared context and coordination updates across agents. Third, loop-aware radio resource control assigns radio resources, QoS flows, and retransmission behavior according to each message's loop role. These capabilities motivate edge-cloud collaborative networking, over-the-air aggregation and federated multi-agent learning, and token- or task-oriented communications.

\begin{itemize}
    \item \textbf{Edge-cloud collaborative networking and service migration.} Edge-cloud collaborative networking supports radio-edge service continuity when agent execution spans user equipment, access points, edge servers, and clouds. It places reasoning, memory, tool sessions, and inference state near the serving wireless link. During handover or edge relocation, service migration transfers the relevant loop state with the edge service, allowing the next decision to continue after mobility.
    \item \textbf{Over-the-air aggregation and federated multi-agent learning.} Over-the-air aggregation and federated multi-agent learning support wireless group-state synchronization when agents update a shared model, policy, or coordination state. Over-the-air aggregation combines local updates through waveform superposition to reduce update latency. Federated multi-agent learning refines the shared model or policy without centralizing raw observations. This treats the wireless channel as a shared-state aggregation medium for multi-agent coordination.
    \item \textbf{Token- and task-oriented communications.} Token- and task-oriented communications support loop-aware radio resource control for agent services. An agent loop may carry token streams, context updates, execution commands, and action results whose value changes with the loop state. Token-oriented transport preserves interactive reasoning, while task-oriented communication links each message to the task state and action it affects. Radio resource control can then prioritize, retransmit, or discard messages according to their loop role.
\end{itemize}

These examples point to a broader research direction in which radio resource control, edge mobility support, and group synchronization are driven by the state evolution, timing constraints, and coordination needs of agent loops.

\section{Conclusion}
Agent-native wireless communications extend AI-native wireless systems by making agent loops part of wireless operation and service delivery. In this view, agents are both communication-system operators and communication-service users. The central design issue is to support the reasoning, action, state, and feedback needed by these loops across radio, edge, and network functions.

This article organized this design space along three architectural components and two functional views. Computing infrastructure provides the hardware platform for agent reasoning and state continuity, O-RAN makes RAN operation observable and programmable to agents, and open interfaces turn agent decisions into authorized network actions with execution evidence. The \emph{agents for communications} view showed how agents generate communication software and coordinate adaptive wireless optimization, while the \emph{communications for agents} view showed how wireless systems support single-agent loops, multi-agent coordination, and agent-loop communications. The discussion above points to four research directions for agent-native wireless systems.

\begin{itemize}
    \item \textbf{Safe multi-agent wireless control.} When scheduling, beam control, slicing, and mobility agents operate over coupled radio and edge resources, an action that improves one control loop may change the interference, load, or latency observed by another. The research issue is coordination before actuation. Agents need a shared view of interference coupling, slice resource budgets, handover timing, and edge load so that incompatible actions can be filtered or sequenced before they reach the network. O-RAN measurements and digital-twin evaluation can provide the common operating state for this conflict-aware control.
    \item \textbf{Mobility-aware agent-loop continuity.} Mobility changes more than the serving link of an agent. It may also move the context cache, task memory, tool session, and edge-side inference state used by the next decision. Agent-native mobility support should therefore treat handover and edge relocation as loop-state transfer procedures, where mobility prediction triggers proactive state placement and the new serving edge resumes the loop without losing decision history.
    \item \textbf{Agent-loop metrics and testbeds.} Conventional wireless metrics reveal link performance, while model metrics reveal inference quality. Agent-native evaluation should instead measure whether the loop remains useful after communication, control, and mobility events. Radio-edge-agent testbeds should record whether context arrives before its decision window expires, whether an action command is executed within its valid interval, whether feedback is matched to the correct action and state version, and whether multiple agents complete the task with consistent shared state.
    \item \textbf{Agent-loop interface standardization.} Even when an agent makes a valid decision, the network still needs to interpret it as a bounded wireless operation. Standardized agent-loop interfaces should define the action schema, loop identity, state version, execution evidence, and rollback support associated with each agent action. These semantics allow reasoning outputs to become executable, auditable, and recoverable network operations across RAN, edge, and service platforms.
\end{itemize}

These directions show that next-generation wireless systems need to support intelligent devices as well as the stateful, verifiable, and coordinated operation of intelligent agents.

\end{document}